\documentclass[conference,10pt]{IEEEtran}
\usepackage{cite}
\usepackage{graphicx}
\usepackage{psfrag}
\usepackage{url}
\usepackage{amsmath}
\usepackage{array}
\usepackage{amssymb}
\usepackage{mathtools}
\usepackage{amsfonts}
\usepackage{graphicx}
\usepackage{epstopdf}
\usepackage{algorithm}
\usepackage{algorithmic}

\usepackage{setspace}
\usepackage{amscd}
\usepackage{mathrsfs}
\usepackage{epsfig}
\usepackage{color}
\usepackage{textcomp}
\usepackage{tikz}
\usetikzlibrary{shapes,snakes,plotmarks}

\usepackage{caption}
\usepackage{subcaption}

\IEEEoverridecommandlockouts

\makeatletter
\newcommand\fs@spaceruled{\def\@fs@cfont{\bfseries}\let\@fs@capt\floatc@ruled
  \def\@fs@pre{\vspace{0.5\baselineskip}\hrule height.7pt depth0pt \kern2pt}%
  \def\@fs@post{\kern2pt\hrule\relax}%
  \def\@fs@mid{\kern2pt\hrule\kern2pt}%
  \let\@fs@iftopcapt\iftrue}
\makeatother

\title{RadioWeaves for Extreme Spatial Multiplexing in Indoor Environments}

\begin{document}
\author{\IEEEauthorblockN{Unnikrishnan Kunnath Ganesan, Emil Bj\"ornson and Erik G. Larsson}
		\IEEEauthorblockA{Department of Electrical Engineering (ISY)\\
		Link\"oping University, Link\"oping, SE-581 83, Sweden.\\
		Emails: \{unnikrishnan.kunnath.ganesan, emil.bjornson, erik.g.larsson\}@liu.se}
\thanks{This work is supported in part by ELLIIT and in part by  Swedish Research Council (VR).}
}
\maketitle
\thispagestyle{empty}

\begin{abstract}
With the advances in virtual and augmented reality, gaming applications, and entertainment, certain indoor scenarios will require vastly higher capacity than what can be delivered by 5G. In this paper, we focus on massive MIMO for indoor environments. We provide a case study of the distributed deployment of the antenna elements over the walls of a room and not restricting the antenna separation to be half the wavelength. This is a new paradigm of massive MIMO antenna deployment, introduced in \cite{van2019radioweaves} under the name RadioWeaves. We investigate different antenna deployment scenarios in line of sight communication. We observe that the RadioWeaves deployment can spatially separate users much better than a conventional co-located deployment, which outweighs the losses caused by grating lobes and thus saves a lot on transmit power. Through simulations, we show that the RadioWeaves technology can provide high rates to multiple users by spending very little power at the transmitter compared to a co-located deployment.
\end{abstract}

\begin{IEEEkeywords} 
RadioWeaves, massive MIMO, zero forcing, radio stripes, patch antennas.
\end{IEEEkeywords}

\section{Introduction}
\label{sec:Introduction}
Massive MIMO introduced in \cite{marzetta2010noncooperative} is a promising technology for 5G and future wireless networks. During the last decade, the demand for data traffic has grown with a growing consensus that wireless systems should support three generic
services namely enhanced mobile broadband (eMBB), massive machine-type communications (mMTC), and ultra-reliable low latency communications (URLLC). Service heterogeneity can be accommodated by massive MIMO and has been studied in great detail \cite{bana2019massive}. To support a large number of users and to provide high per-user data rates, one primary approach is the densification of the network infrastructure or deploying small cells. However, as we densify the network, the performance is limited by inter-cell interference. To overcome this limitation, cellular networks without cell boundaries, known as cell-free massive MIMO \cite{ngo2017cell}, were studied and shown to be a promising solution. The authors of \cite{interdonato2019ubiquitous} came up with the idea of deploying antenna by meters, known as radio stripes, as a practical implementation of cell-free massive MIMO. 

The RadioWeaves technology, introduced in \cite{van2019radioweaves}, presents a new wireless networking infrastructure where antennas and underlying electronics are weaved into conventional buildings and objects. The technology leverages on distributing the
antenna arrays over large surface areas and is believed to have unprecedented capabilities for cell-free infrastructure offering service connectivity, energy efficiency, reliability, and low latency.

With the advances in virtual and augmented reality, there has been a huge demand for data traffic for gaming applications and entertainment in indoor scenarios like offices and malls, and require vastly higher capacity than what can be delivered by 5G. Base stations deployed in the outside premises may not be able to support such a huge traffic demand due to shadowing and certain building constructions make it difficult for the signals to penetrate. Also, if the base station is far away, it is harder to spatially separate many users in a room. Hence, in this paper, we do a case study about different antenna deployment topologies for indoor environments. As the antennas are deployed indoors, we assume that there are direct line-of-sight paths from all the antennas to each user. We consider two types of antenna deployment for the indoor scenarios in this paper:
\begin{enumerate}
\item Co-located deployment where the antennas are separated by half the wavelength.
\item RadioWeaves deployment where antenna arrays are distributed over large surfaces within a room, where the antenna separation is not restricted to half the wavelength but can be larger.
\end{enumerate}

We address the following questions:
\begin{enumerate}
\item  If we are supposed to deploy a certain number of antennas in a room, is it preferable to have a co-located array or RadioWeaves array deployment?
\item In case of a RadioWeaves deployment, is it preferable to deploy a linear array (1 dimension) or a grid-array (2 dimensions) over the wall?
\item Should we deploy on a single wall or multiple walls?
\end{enumerate}

We observe that the RadioWeave deployment can spatially separate the users much better than the co-located deployment. The spatial separation of users allows the transmitter to reduce the total required power to a great extent. The spatial separation of users is higher when the antennas are distributed over all the walls of the room. The authors of \cite{de2020csi} studied the line-of sight channel state information measured by 64 antennas which were deployed as uniform rectangular array, uniform linear array and distributed over line, with four single-antenna users positioned in an office. In this paper, we also consider how the performance is affected by errors in channel estimates when a large number of users are considered. Through simulations, we show that, RadioWeaves technology can provide high rates to multiple users by spending very little power compared to the co-located deployment.

\textbf{\textit{Notations:}} Bold, lowercase letters are used to denote vectors and bold, uppercase letters are used to denote matrices. $\mathbb{C}$ denote the set of complex numbers. For a matrix $\mathbf{A}$, $\mathbf{A}^*$, $\mathbf{A}^T$ and $\mathbf{A}^{H}$ denotes conjugate, transpose and conjugate transpose of the matrix $\mathbf{A}$ respectively. $\mathcal{CN}(0,\sigma^2)$ denotes a circularly symmetric complex Gaussian random variable with zero mean and variance equal to $\sigma^2$. $\text{Tr}(\mathbf{A})$ denotes trace of $\mathbf{A}$. Identity matrix is denoted by $ \mathbf{I} $. 

\section{Preliminaries}
\label{sec:PrelimsChannelModel}
In this section, we outline certain preliminaries related to conversion of the coordinate system from cartesian to polar coordinates, free space signal propagation and microstrip patch antennas which help in further understanding of the paper. 
\subsection{Coordinate System}
Consider a point $ (x, y, z) $ in the cartesian coordinate system. The spherical coordinates $ (r, \theta, \varphi) $, where $ r $ is the distance to the point from the origin, $ \theta\in [0,\pi] $ is the elevation angle and $\varphi\in [-\pi, \pi) $ is the azimuth angle, can be obtained from $ (x, y, z) $ as
\begin{align}
r & = \sqrt{x^2+y^2+z^2}, \\
\cos \theta & = \left(\frac{z}{r}\right), \\
\tan \varphi & = \left(\frac{y}{x}\right).
\end{align}

\subsection{Free Space Signal Propagation}
Let $ P_{rad} $ be the power radiated by the transmitter located at the origin. The received power by an isotropic antenna at location $ (r, \theta, \varphi) $ operating at wavelength $\lambda$ is given by 
\begin{equation}
\label{eqn:RxPower}
P_{rx} = \frac{P_{rad}}{4\pi} \ G(\theta,\varphi) \ \frac{1}{r^2} \ \frac{\lambda^2}{4\pi} ,
\end{equation}
where $G(\theta,\varphi)$ is the directional power gain of the transmit antenna which can be computed as
\begin{equation}
G(\theta,\varphi) = \frac{4\pi \  U(\theta,\varphi)}{\int \int U(\theta,\varphi) \sin(\theta) d\theta d\varphi} = \alpha^2 \ U(\theta,\varphi),
\end{equation}
and $U(\theta,\varphi)$ is the normalized directional radiation intensity. For an isotropic transmitting antenna, $U(\theta,\varphi) = \frac{1}{4\pi}$ and hence $G(\theta,\varphi)=1$ and when $G(\theta,\varphi)$ is a constant, (\ref{eqn:RxPower}) gives Friis' free space equation \cite{friis1946note}.

\subsection{Microstrip Patch Antenna}
\label{ssec:AntennaDesign}
For indoor scenarios, it is preferable to deploy directional antennas as there is no need to send any power into the walls. Thus, we consider each transmitter antenna to be a rectangular microstrip patch antenna \cite[Ch. 14]{balanis2016antenna} with dimensions $ h $, $ L $, and $ W $, which stand for the height, length, and width of the patch antenna, respectively. In this subsection, we give the design procedure of the patch antenna outlined in \cite[Ch. 14]{balanis2016antenna}. Let $ \varepsilon_r $ and $ f $ be the dielectric constant of the substrate and resonant frequency respectively. Let $ c $ be the speed of light. For a given $ h $, the design of the patch is given as follows
\begin{align}
W & = \frac{c}{2f}\sqrt{\frac{2}{\varepsilon_r+1}} \\
L & = \frac{c}{2f\sqrt{\varepsilon_{\textrm{reff}}}}-2\Delta L
\end{align}
where $\varepsilon_{\textrm{reff}}$ and $\Delta L$ are obtained as 
\begin{align}
\varepsilon_{\textrm{reff}} & = \frac{\varepsilon_r+1}{2} + \frac{\varepsilon_r-1}{2}\left( 1+12\frac{h}{W} \right)^{-1/2} \\
\frac{\Delta L}{h} & = 0.412\frac{(\varepsilon_{\textrm{reff}}+0.3)(\frac{W}{h}+0.264)}{(\varepsilon_{\textrm{reff}}-0.258)(\frac{W}{h}+0.8)}.
\end{align}

\subsection{Line of Sight Channel Model for Micro strip Patch Antenna}
From (\ref{eqn:RxPower}), the line-of-sight channel gain $g$ between the transmitter antenna and receiver antenna can be written as 
\begin{equation}
\lvert g \rvert^2 = \left(\frac{\lambda}{4\pi r}\right)^2   G(\theta,\varphi). 
\end{equation}
In this paper, we consider the radiating elements are vertically polarized and assume that there is no polarization losses. For a microstrip patch antenna, $g$ can be modeled as \cite{balanis2016antenna},
\begin{equation}
\label{eqn:LosChannel}
g = \frac{\alpha\lambda}{4\pi r}e^{-j\frac{2\pi r}{\lambda}}
\sin(\theta) \frac{\sin (X) }{X} \frac{\sin (Z)}{Z}
\end{equation}
where 
\begin{align}
X & =\frac{\pi h}{\lambda}\sin(\theta)\cos(\varphi) \\
Z & =\frac{\pi W}{\lambda}\cos(\theta) \\
\alpha^2 & = \frac{4\pi}{\displaystyle\int_{\theta=0}^{\pi} \displaystyle\int_{\varphi=-\frac{\pi}{2}}^{\frac{\pi}{2}} \left( \frac{\sin (X)}{X} \frac{\sin (Z)}{Z} \right)^2 \sin^3(\theta) d\theta d\varphi}.
\end{align}

\section{System Model and Analysis}
In this section, we introduce the system model and study the performance of the system with and without channel state information at the base station.
\subsection{System Model}
Let us consider an indoor environment where $ M $ patch antennas are serving $ K $ single-antenna users. Let $ (r_{mk} , \theta_{mk} , \varphi_{mk} ) $ be the relative position of receiver $ k $ with respect to transmitter $ m $ in the spherical coordinate system. The line-of-sight channel between the $ m^{th} $ transmit antenna and the $ k^{th} $ receiver is represented by $ g_{mk} $ and is given by (\ref{eqn:LosChannel}) with position $ (r_{mk} , \theta_{mk} , \varphi_{mk} ) $. Let $ \mathbf{G} \in \mathbb{C}^{ M\times K} $ be the channel matrix between the $ M $ transmit antennas and the $ K $ users defined by

\begin{equation}
\label{eqn:ChannelMatrix}
\mathbf{G} = \left[ \begin{matrix}
g_{11} & g_{12} & \dots & g_{1K} \\
g_{21} & g_{22} & \dots & g_{2K} \\
\vdots & \vdots & \ddots & \vdots \\
g_{M1} & g_{M2} & \dots & g_{MK}
\end{matrix}  \right] .
\end{equation}

We consider that there are no scatterers in the system, hence, we have only a direct line-of-sight path between the base station and users. As we are not restricting the antenna element seperation be to half the wavelength, we assume that there is no coupling between the transmitting antennas at the base station. 

\subsection{Line-of-Sight Channel Estimation}
In this subsection, we consider the estimation of the line-of-sight channel. As the channel is deterministic, we estimate the channel in every transmission block or frame. Let $\tau_c$ denote the number of symbols in a transmission block. We assume that all the users are assigned mutually orthogonal pilot sequences of length $\tau_p\geq K$ and $\tau_p\ll\tau_c$. Let $\boldsymbol{\phi}_k \in\mathbb{C}^{\tau_p\times1}$ be the pilot sequence associated with user $k$, which is the $k^{th}$ column of the $\tau_p\times K$ unitary matrix $\boldsymbol{\Phi}$ and $\boldsymbol{\Phi}^H\boldsymbol{\Phi} = \mathbf{I}$. Each user transmits $\sqrt{\tau_p}\boldsymbol{\phi}_k^H$ over $\tau_p$ symbols and the signals collectively received at the base station terminals $\mathbf{Y}_p\in\mathbb{C}^{M\times \tau_p}$ can be written as 
\begin{align} 
\mathbf{Y}_p = \sqrt{\rho_{\text{ul}}\tau_p}\mathbf{G} \boldsymbol{\Phi}^H + \mathbf{W}_p,
\end{align}
where $\rho_{\text{ul}}$ is a scalar to scale the uplink transmit power and $\mathbf{W}_p$ is the noise matrix with i.i.d elements $\mathcal{CN}(0,\sigma^2)$. De-spreading operation is performed at the base station on the received signal $\mathbf{Y}_p$ and is given by
\begin{align} 
\mathbf{Y}_p' = \mathbf{Y}_p\boldsymbol{\Phi} = \sqrt{\rho_{\text{ul}}\tau_p}\mathbf{G} + \mathbf{W}_p'. 
\end{align}
Since $\boldsymbol{\Phi}$ is a unitary matrix, the elements of $\mathbf{W}_p'=\mathbf{W}_p\boldsymbol{\Phi}$ are i.i.d. $\mathcal{CN}(0,\sigma^2)$. 
The channel is not fading, so there is no statistics and hence we use the least squares method for estimating the channel and is given by 
\begin{align}
\label{eqn:LSEst}
\hat{g}_{mk} = \frac{[\mathbf{Y}_p']_{mk}}{\sqrt{\rho_{\text{ul}}\tau_p}}
\end{align}
where $[\mathbf{Y}_p']_{mk}$ is the $(m,k)^{th}$ element of $\mathbf{Y}_p'$. Let $\mathbf{\hat{G}} \in \mathbb{C}^{M\times K}$ be the estimated channel matrix defined analogous to (\ref{eqn:ChannelMatrix}) using the channel estimates in (\ref{eqn:LSEst}). 

\subsection{Power Required for Zero Forcing Operation}
In this paper, we consider zero forcing precoding \cite{wiesel2008zero} at the base station in the  downlink in order to give all the users similar rates. In this subsection, we compute the power spent at the base station for zero forcing operation. Let $ q_k \sim \mathcal{CN} (0, 1) $ be the downlink signal intended for user $ k $ and let $ \mathbf{q} = [q_1 \ q_2 \ \dots \ q_K ]^T \in \mathbb{C}^{ K\times1} $. The zero forcing precoded signal $\mathbf{x}\in\mathbb{C}^{M\times 1}$ is given by
\begin{equation}
\mathbf{x} = \mathbf{\hat{G}}^*(\mathbf{\hat{G}}^T\mathbf{\hat{G}}^*)^{-1}\mathbf{q}.
\end{equation}

The power spent at the transmitter to achieve the zero-forced signal is given by
\begin{align}
\label{eqn:TotalPowerRequired}
P & =  \lVert \mathbf{x} \rVert ^2  = \text{Tr}((\mathbf{\hat{G}}^T\mathbf{\hat{G}}^*)^{-1}).
\end{align}

%


\subsection{Downlink Spectral Efficiency}
From the transmission block, $\tau_c-\tau_p$ symbols are used for downlink. We consider zero forcing precoding using the channel estimates for downlink, and the signal from base station to all users is given by 
\begin{equation}
\mathbf{x} = \mathbf{A}\mathbf{q},
\end{equation}
where $\mathbf{A}=\frac{1}{\sqrt{\text{Tr}((\mathbf{\hat{G}}^T\mathbf{\hat{G}}^*)^{-1})}}\mathbf{\hat{G}}^*(\mathbf{\hat{G}}^T\mathbf{\hat{G}}^*)^{-1} = [\mathbf{a}_1 \ \mathbf{a}_2 \ \dots \ \mathbf{a}_K]$ such that $\mathbb{E}\{\lVert\mathbf{x}\rVert^2 \}=1$. The collective signals received at the user terminals is given by
\begin{align}
\mathbf{y} & = \sqrt{\rho_{\text{dl}}}\mathbf{G}^T\mathbf{x} + \mathbf{w},
\end{align}
where $\rho_{\text{dl}}$ is the total downlink power and $ \mathbf{w}=[w_1 \ w_2 \ \dots \ w_K]^T $ is the noise vector whose elements are independent and identically distributed $ \mathcal{CN} (0, \sigma^2 ) $ and $ \sigma^2 $ is the noise power. Now, the signal received at each user $k$, can be written as 
\begin{equation}
\label{eqn:yk}
y_k = \sqrt{\rho_{\text{dl}}}\sum_{i=1}^{K}\mathbf{g}_k^T\mathbf{a}_iq_i + w_k,
\end{equation}
which can be equivalently written as 
\begin{equation}
\label{eqn:ykmod}
\begin{aligned}
y_k = & \sqrt{\rho_{\text{dl}}} \mathbb{E}\{\mathbf{g}_k^T\mathbf{a}_k\}q_k + \sqrt{\rho_{\text{dl}}}(\mathbf{g}_k^T\mathbf{a}_k - \mathbb{E}\{\mathbf{g}_k^T\mathbf{a}_k\})q_k \\
& + \sqrt{\rho_{\text{dl}}}\sum_{i=1,i\neq k}^{K}\mathbf{g}_k^T\mathbf{a}_iq_i + w_k.
\end{aligned}
\end{equation}
The terms in (\ref{eqn:ykmod}) are uncorrelated, and hence the signal-to-interference-noise-ratio (SINR) at user $k$ \cite[Ch. 2, Sec. 2.3.2]{marzetta2016fundamentals} can be written as 
\begin{equation}
\label{eqn:SINR}
\text{SINR}_k = \frac{\rho_{\text{dl}}\lvert\mathbb{E}\{\mathbf{g}_k^T\mathbf{a}_k\}\rvert^2 }{\sigma^2 + \rho_{\text{dl}}\sum_{i=1,i\neq k}^{K}\mathbb{E}\{\lvert\mathbf{g}_k^T\mathbf{a}_i\rvert^2\} + \rho_{\text{dl}} \text{var}\{\mathbf{g}_k^T\mathbf{a}_k\}}.
\end{equation}
In (\ref{eqn:SINR}), all the randomness is in the channel estimation errors and the ergodicity is with respect to the channel estimation errors. Thus, the rate which user $k$ can achieve is given by
\begin{align}
 R_k = \left(1-\frac{\tau_p}{\tau_c}\right)\log(1+\text{SINR}_k).
\end{align}

\begin{table}[t]
	\caption{Simulation Parameters}
	\label{table:SimulationSetup}
	\begin{center}
		\begin{tabular}{ |l| c| }
			\hline
			Frequency of operation, $f$ & $ 2 $ GHz  \\ \hline 
			Wavelength, $\lambda$ & $ 15 $ cm  \\  \hline   
			Room size & $ 40 $ m x $ 40 $ m x $ 10 $ m     \\ \hline
			Number of transmit antennas, $M$ & $ 512 $ \\ \hline
			Number of users, $K$ & $ 200 $ \\ \hline
			Noise figure, $F$ & $ 9 $ dB \\ \hline
			Bandwidth, $B$ & $ 20 $ MHz \\ \hline
			Noise power, $\sigma^2$ & $-92$ dBm \\ \hline
			Dielectric constant, $\varepsilon_r$ & $ 10.2 $ \\ \hline
			Height of patch antenna, $h$ & $ 0.1588 $ cm \\ \hline
		\end{tabular}
	\end{center}
\end{table}

\section{Case Study}
\label{sec:Simualtions}
\begin{figure}[t]
	\includegraphics[scale=0.6]{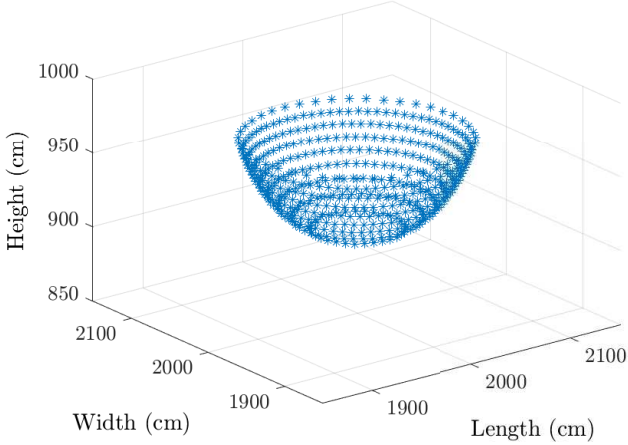}
	\caption{Co-located massive MIMO deployment in the shape of
		a candelabrum over the ceiling.}
	\label{fig:candelabrum}
\end{figure}

We consider a simulation scenario consisting of an indoor room of size $ 40 \times 40 \times 10 \ \text{m}^3$. The simulation parameters are summarized in Table \ref{table:SimulationSetup}. We consider different topologies for the deployment of antennas. For the co-located reference case, we consider a candelabrum type deployment on the ceiling of the room and is shown in Fig. \ref{fig:candelabrum}. This shape guarantees that there are antennas pointing towards every corner of the room. 

\begin{figure}[t]
	\includegraphics[scale=0.5]{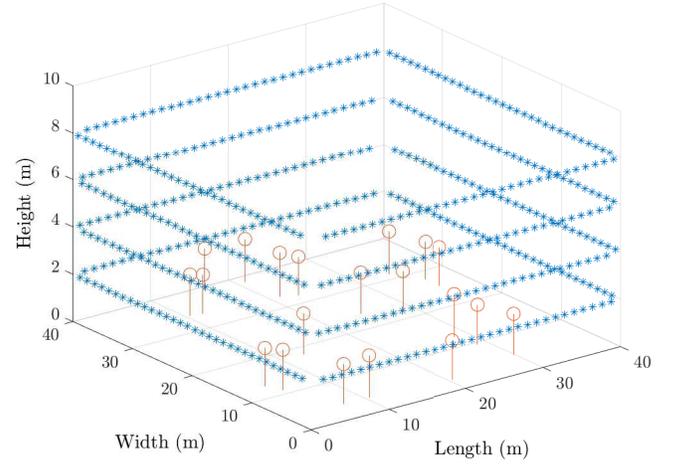}
	\caption{Quadruple strip deployed on all four walls of the room}
	\label{fig:FourStripe4Wall}
\end{figure}

For RadioWeaves deployment we consider placing the antennas and underlying electronics circuitry over a stripe (also known as radio stripe \cite{interdonato2019ubiquitous}) and the following deployment topologies are considered: 
\begin{enumerate}
\item Single strip on 1 wall: The antennas are deployed horizontally over a single wall with $ \frac{\lambda}{2} $ spacing.
\item Single strip on 4 walls: The antennas are deployed horizontally over all walls with $ 2\lambda $ spacing.
\item Double strip on 1 wall: The antennas are deployed horizontally as two separate stripes with vertical spacing of 2 m over a single wall with $\lambda$ spacing.
\item Double strip on 4 walls: The antennas are deployed horizontally as two separate stripes with vertical spacing of 2 m over all walls with $ 4\lambda $ spacing.
\item Quadruple strip on 1 wall: The antennas are deployed horizontally as four separate stripes with vertical spacing of 2 m over a single wall with $ 2\lambda $ spacing.
\item Quadruple strip on 4 walls: The antennas are deployed horizontally as four separate stripes with vertical spacing of 2 m over all walls with $ 8\lambda $ spacing. A typical deployment scenario is shown in Fig. \ref{fig:FourStripe4Wall}.
\end{enumerate}

\begin{figure}[t]
	\includegraphics[scale=0.47]{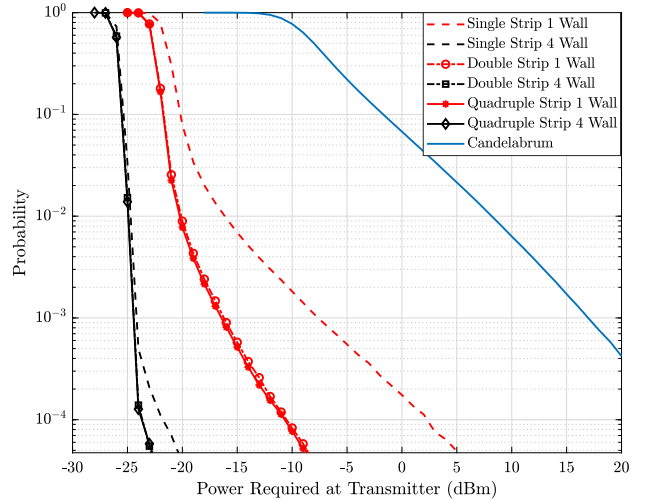}
	\caption{Probability of required transmit power to achieve minimum spectral efficiency of $ 4 $ bit/s/Hz per user.}
	\label{fig:PowerComparison}
\end{figure}

We convey two things in this paper through simulations. First, we show that the total downlink power required when we deploy RadioWeave technology is very low compared to the co-located deployment when zero-forcing precoding is used at the base station. As the channel estimation quality depends on the uplink power, for the comparison of required downlink power among different topologies, we consider an uplink power of 0dBm, such that the channel estimation errors are negligible. From (\ref{eqn:TotalPowerRequired}), we can see that the required transmit power $ P $ depends on the channel matrix $\mathbf{\hat{G}}$ and hence is random depending on user locations. For simulations, we consider $ 10^6 $ random user locations and as a metric for comparison, we plot the probability of required transmit power, i.e., $ \text{Pr}\{P \geq \text{power}\} $ to achieve a minimum downlink spectral efficiency requirement of $ 4 $ bit/s/Hz per user. Fig. \ref{fig:PowerComparison} shows the probability of required transmit power for different antenna deployment topologies keeping the total number of transmit antennas the same. It can be seen that the power required in the co-located deployment is higher compared to all the RadioWeaves deployment scenarios considered in the paper. From Fig. \ref{fig:candelabrum}, it can be observed that not all antennas will be able to serve each user, as some antennas point in the wrong direction. Hence, in the co-located case, we are not able to achieve the full multiple antenna power gain. The transmit power in the RadioWeaves deployment topologies considered is lower since the users can be spatially separated as all the transmit antennas can see all users, thus achieving a higher power gain from coherent transmission. Also, we observe that, if the antennas are divided over all four walls, the required transmit power is much less compared to keeping all the antennas on the same wall. This means that the capability of the antenna array to spatially resolve each user improves as the antennas are spread over all the walls. Thus, for a RadioWeaves deployment, distributing the antennas over all the walls of the room is preferred. Also, the power savings or spatial resolution capability of the array is negligible when the array is deployed vertically over the walls. From the plot, it can be seen that RadioWeaves with deploying antennas over all the four walls of the indoor building can provide a minimum downlink rate of $ 4 $ bit/s/Hz for 200 users with a total required transmit power of $ -23 $ dBm. 


\begin{figure}[t]
	\includegraphics[scale=0.55]{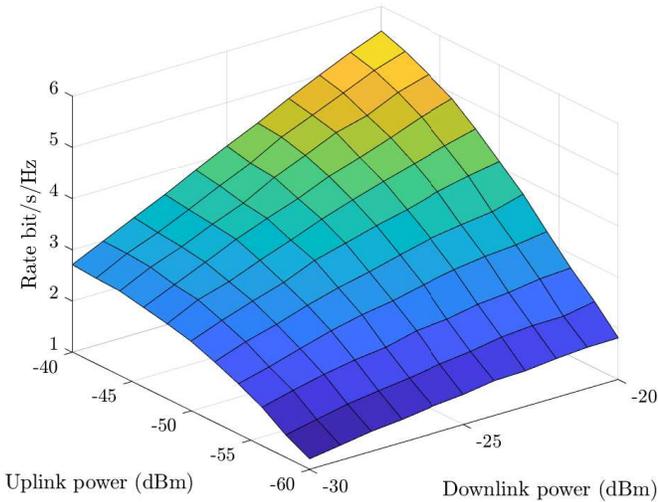}
	\caption{99.9\% likely achievable rate per user in a “double strip on 4 walls” configuration under different uplink downlink power levels.}
	\label{fig:SpectralEfficiency_UpDownlinkPower}	
\end{figure}

In the antenna array gain processing literature, it is known that the larger than half the wavelength antenna element spacing deployments are not desirable due to the grating lobes. This is true when we consider an equal aperture array and compare the performance between antenna elements spaced half the wavelength and antenna elements spaced more than half the wavelength by removing antenna elements. In the latter case, we get grating lobes which reduces the spatial resolution of users. However, if we put the removed antenna elements at another location or on a different wall, thereby allowing a larger array, we get an improvement in spatial resolution that outweigh the losses caused by the grating lobes. As the antenna separation is increased, the total angular window covered when pointing the main lobe into a particular direction remains more or less unchanged even though the lobes get narrower and starts getting grating lobes. Thus when all the antenna elements are used and deployed all around the indoor space, the grating lobes do not effect whatsoever on the performance. 

Secondly, we show that, with low power levels, high rates are achievable with RadioWeaves technology. Fig. \ref{fig:SpectralEfficiency_UpDownlinkPower} shows the 99.9\% likely rates users can achieve with different uplink and downlink power levels. From (\ref{eqn:LSEst}), the channel estimates depend on the uplink power. At low uplink power levels, the channel estimates are not good and which in turn reduces the downlink rate. As the uplink power is increased, first we see a gradual improvement in rates after which the channel estimates quality becomes close to perfect channel state and the rate saturates. As the downlink power is increased, the rates increases gradually. The plot shows that the RadioWeave deployment can provide high rates to the users by spending very little power.

\section{Conclusion}
In this paper, we did a case study about different antenna deployment topologies in an indoor environment when there are only direct line-of-sight paths between the transceivers. The RadioWeave deployment of antennas is shown to be a promising approach over the co-located deployment of antennas, which reduces the transmit power and provide a high-quality service to the users. From the case study, we see that the grating lobes occurring due to larger antenna spacing does not affect the performance when all elements are used over a larger aperture. Simulation results show that deploying grid-array over multiple walls without any restriction on antenna spacing being half the wavelength, is preferable.  

\bibliographystyle{IEEEtran}
\bibliography{references}

\begin{thebibliography}{10}
\providecommand{\url}[1]{#1}
\csname url@samestyle\endcsname
\providecommand{\newblock}{\relax}
\providecommand{\bibinfo}[2]{#2}
\providecommand{\BIBentrySTDinterwordspacing}{\spaceskip=0pt\relax}
\providecommand{\BIBentryALTinterwordstretchfactor}{4}
\providecommand{\BIBentryALTinterwordspacing}{\spaceskip=\fontdimen2\font plus
\BIBentryALTinterwordstretchfactor\fontdimen3\font minus
  \fontdimen4\font\relax}
\providecommand{\BIBforeignlanguage}[2]{{%
\expandafter\ifx\csname l@#1\endcsname\relax
\typeout{** WARNING: IEEEtran.bst: No hyphenation pattern has been}%
\typeout{** loaded for the language `#1'. Using the pattern for}%
\typeout{** the default language instead.}%
\else
\language=\csname l@#1\endcsname
\fi
#2}}
\providecommand{\BIBdecl}{\relax}
\BIBdecl

\bibitem{van2019radioweaves}
L.~Van~der Perre, E.~G. Larsson, F.~Tufvesson, L.~De~Strycker, E.~Bj{\"o}rnson,
  and O.~Edfors, ``Radioweaves for efficient connectivity: analysis and impact
  of constraints in actual deployments,'' in \emph{2019 53rd Asilomar
  Conference on Signals, Systems, and Computers}.\hskip 1em plus 0.5em minus
  0.4em\relax IEEE, 2019, pp. 15--22.

\bibitem{marzetta2010noncooperative}
T.~L. Marzetta, ``{Noncooperative cellular wireless with unlimited numbers of
  base station antennas},'' \emph{IEEE Trans. on Wireless Commun.}, vol.~9,
  no.~11, pp. 3590--3600, 2010.

\bibitem{bana2019massive}
A.-S. Bana, E.~De~Carvalho, B.~Soret, T.~Abr{\~a}o, J.~C. Marinello, E.~G.
  Larsson, and P.~Popovski, ``{Massive MIMO for Internet of Things (IoT)
  Connectivity},'' \emph{Physical Communication}, vol.~37, pp. 1--17, 2019.

\bibitem{ngo2017cell}
H.~Q. Ngo, A.~Ashikhmin, H.~Yang, E.~G. Larsson, and T.~L. Marzetta,
  ``{Cell-free massive MIMO versus small cells},'' \emph{IEEE Transactions on
  Wireless Communications}, vol.~16, no.~3, pp. 1834--1850, 2017.

\bibitem{interdonato2019ubiquitous}
G.~Interdonato, E.~Bj{\"o}rnson, H.~Q. Ngo, P.~Frenger, and E.~G. Larsson,
  ``{Ubiquitous cell-free massive MIMO communications},'' \emph{EURASIP Journal
  on Wireless Communications and Networking}, vol. 2019, no.~1, p. 197, 2019.

\bibitem{de2020csi}
S.~De~Bast, A.~P. Guevara, and S.~Pollin, ``{CSI-based positioning in massive
  MIMO systems using convolutional neural networks},'' in \emph{2020 IEEE 91st
  Vehicular Technology Conference (VTC2020-Spring)}.\hskip 1em plus 0.5em minus
  0.4em\relax IEEE, 2020, pp. 1--5.

\bibitem{friis1946note}
H.~T. Friis, ``A note on a simple transmission formula,'' \emph{Proceedings of
  the IRE}, vol.~34, no.~5, pp. 254--256, 1946.

\bibitem{balanis2016antenna}
C.~A. Balanis, \emph{{Antenna Theory: Analysis and Design}}.\hskip 1em plus
  0.5em minus 0.4em\relax John Wiley \& sons, 2016.

\bibitem{wiesel2008zero}
A.~Wiesel, Y.~C. Eldar, and S.~Shamai, ``{Zero-forcing precoding and
  generalized inverses},'' \emph{IEEE Transactions on Signal Processing},
  vol.~56, no.~9, pp. 4409--4418, 2008.

\bibitem{marzetta2016fundamentals}
T.~L. Marzetta, E.~G. Larsson, H.~Yang, and H.~Q. Ngo, \emph{Fundamentals of
  massive MIMO}.\hskip 1em plus 0.5em minus 0.4em\relax Cambridge University
  Press, 2016.

\end{thebibliography}

\end{document}